\documentclass[aps,prd,twocolumn,groupedaddress,showpacs]{revtex4}
\usepackage{amssymb}
\usepackage{graphicx}
\usepackage{bm}
\usepackage{amsmath}

\newcommand{\rp}{r_{+}}

\newcommand{\LF}{{\cal L}(F)}
\newcommand{\dO}{d\Omega^{2}}
\newcommand{\ro}{r_{0}}

\begin{document}


\title{Extremal limit of the regular charged black holes 
in nonlinear electrodynamics}
\author{Jerzy Matyjasek}
\email{matyjase@tytan.umcs.lublin.pl}
\affiliation{Institute of Physics, 
Maria Curie-Sk\l odowska University\\
pl. Marii Curie-Sk\l odowskiej 1, 
20-031 Lublin, Poland}

\date{\today}

\begin{abstract}
The near horizon limit of the extreme nonlinear black hole 
is investigated. It is shown that 
resulting geometry belongs to the 
${\rm AdS}_{2}\times {\rm S}^{2}$ class with different modules
of curvatures of subspaces and could be described in terms of
the Lambert functions. It is demonstrated 
that the considered class of Lagrangians does not admit 
solutions of the Bertotti-Robinson type.
\end{abstract}


\maketitle

Still growing interest in the extremal  black holes that is seen
recently is motivated by their unusual and not fully understood
nature. The problems of entropy,  semiclasical configurations,
interactions with matter, or information paradox -- to name a few --
have not been resolved yet. Apart from their global structure and
behaviour, the near-horizon region is also of interest. Indeed,
applying appropriate limiting procedure in the geometry of the extreme
and near extreme black holes one can generate new exact
solutions~\cite{Ginsparg,Oleg:1997,Caldarelli,Dadhich,Dias,Cardoso}.

On the other hand, of the equal importance are the questions of the
nature of singularities that reside, hidden to external observer, in the
centers of most black holes. According to the widespread opinion,
singularities that plague such models are symptoms of illness of the
theory rather than its health and the role of the regular solutions in
our understanding of the black hole physics could not be overestimated. 
One of the methods which
can be used in constructions of non-singular models is replacing black
hole interior by a regular core. This idea appeared almost forty years
ago, in mid sixties~\cite{Sakharov,Gliner,Bardeen:68} and is actively
investigated today (see Ref.~\cite{Irina1} for a discussion and a
comprehensive list of references). The exact analytical solution for a
specific sources has been constructed in
Refs.~\cite{Irina2,Borde,Mars}.

Among various regular models known to date, especially intriguing are
the solutions to the coupled equations of nonlinear electrodynamics
and general relativity found by Ay\'on--Beato and
Garc\'{\i}a~\cite{ABG} and  by Bronnikov~\cite{Bronnikov:PRD}. The
latter describes magnetically charged black hole, and provides an
interesting example of the system that could be both regular and
extremal. Moreover, its simplicity allows exact treatment as the
location of the horizons can be expressed in terms of the Lambert
functions~\cite{Jirinek}.

In this note we shall investigate the extremal magnetically charged
black hole with the special emphasis put on the near horizon geometry
and its relation with the exact solutions of the Einstein field
equations with topology
${\rm AdS}_{2}\times {\rm S}^{2}.$
To begin with, however, we remind a few basic facts.

Let us consider general relativistic formulation of nonlinear
electrodynamics described by a class of a gauge invariant Lagrangians
${\cal L}(F),$ where $F\,=\,F_{ab}F^{ab}$ and
$F_{ab}\,=\,\partial_{a}A_{b}\,-\,\partial_{b}A_{a}.$ According to the
well-known theorem~\cite{Bronnikov:AP,Bronnikov:PRL} if ${\cal L}(F)$
has a Maxwell asymptotic for weak fields, i.e., ${\cal L}(F)\,\sim\,F$
and ${\cal L}_{F}\,=\,d{\cal L}/dF\,\to \,{\rm const}$ as $F\,\to 0,$
then any static and spherically symmetric solutions to the coupled
equations of the nonlinear electrodynamics and general relativity with
nonzero electric charge cannot have a regular center. It does not mean
that this no go theorem forbids existence of regular black holes in
general. Simple example of a regular magnetic solution has been
presented by Bronnikov in Ref.~\cite{Bronnikov:PRD}. It bears a formal
resemblance to the solution found earlier by  Ay\'on--Beato and
Garc\'{\i}a~\cite{ABG} describing electrically charged configuration.
It should be noted however that the Ay\'on--Beato and Garc\'{\i}a
solution has been constructed with the use of different Lagrangians
valid in different regions~\cite{Bronnikov:PRD}. Attempts to
circumvent the conditions of the no go theorem have been undertaken by
Burinskii and Hildebrandt~\cite{Burinskii}. They demonstrated that
modifications of the Ay\'on--Beato and Garc\'{\i}a model, in which
electric field does not extend to the central region are in concord
with the no go theorem

The coupled system of equations obtained from the total action
\begin{equation}
S\,=\,\frac{1}{16\pi}\int d^{4}x \sqrt{-g} \left[R\,-\,{\cal L}(F) \right]
                                         \label{action}
\end{equation}
for the static and spherically symmetric configuration 
described by the line element of the form
\begin{equation}
 ds^{2}\,=\,-f(r) dt^{2}\,+\,\frac{1}{f(r)}dr^{2}\,+
 \,r^{2}\dO, 
                                        \label{lineel}
 \end{equation}
 \begin{equation}
 f(r)\,=\,1\,-\,\frac{2 m(r)}{r},
 \end{equation}
with nonlinear
magnetic field reduce to simple quadrature
\begin{equation}
m(r)\,=\,\frac{1}{4}\int  {\cal L}(F)r^{2} dr\,+\,C,
                                         \label{quadrature1}
\end{equation}
where $C$ is an integration constant.
Integrating Eq.~(\ref{quadrature1}) with function ${\cal L}(F)$ given by
 \begin{equation}
 {\cal L}(F)\,=\,F \cosh^{-2}\left[a \left(\frac{F}{2}\right)^{1/4} \right],
                                        \label{lagr}
 \end{equation}
 where $F\,=\,2Q ^{2}/r^{4}$ and $a$ is a free parameter, one obtains 
\begin{equation}
m(r)\,=\,C\,-\,\frac{|Q|^{3/2}}{2a}\tanh\left(\frac{a|Q|^{1/2}}{r} \right).
\end{equation}
Employing boundary condition $m(\infty)\,=\,M$ and setting $a\,=\,|Q|^{3/2}/2M$ yields
\begin{equation}
 f(r)\,=\,1\,-\,\frac{2 M}{r}\left(1\,-\,\tanh\frac{Q^{2}}{2Mr} \right).
                                        \label{fr} 
 \end{equation}
 
\begin{figure}[th]
 \includegraphics[height=6cm]{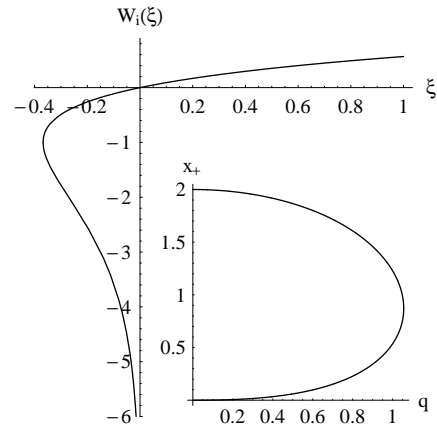}
 \caption{\label{figg} Location of the horizons $x_{+}$ (upper curve)
 and $x_{-}$ (bottom curve) plotted against $q.$ The run of the real branches 
 of the Lambert function $W_{0}(\xi)$ (upper curve) and $W_{-1}(\xi)$ (lower curve)
 is displayed for comparison. The degenerate event horizon at $x_{+}\,=\,2 w_{0}^{1/2}$
 corresponds to the branch point of the Lambert function at $\xi\,=\,-1/e.$}
\end{figure}

 One of the most attractive features of this solution is possibility
 to express the location of the horizons in terms 
 of the Lambert functions, $W_{i}(\xi).$
 The Lambert functions~\cite{Knuth}  defined by means of the general 
 formula
 \begin{equation}
   \exp (W(\xi)) W(\xi)\,=\,\xi,
                                             \label{lamb}
  \end{equation}
  have two real branches, $W_{0}(\xi)$ and $W_{-1}(\xi)$ with a branch
  point $-1/e.$ The run of the real branches of the Lambert functions
  is displayed in Fig.~\ref{figg}. Since the value of the principal
  branch of the Lambert function at $1/e$ plays an important role in
  our considerations we put
  \begin{equation}
  W_{0}(1/e)\,=\,w_{0}.
  \end{equation}

\begin{figure}[th]
 \includegraphics[height=4cm]{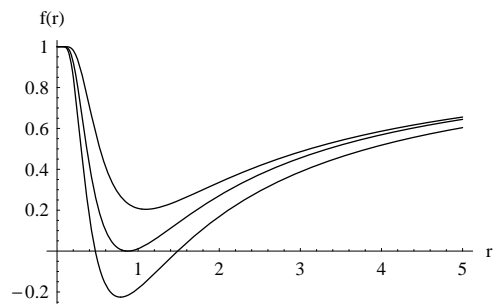}
 \caption{\label{figgg} Qualitative behaviour of the function $f(x)$
 for various values of the parameter $q.$ Top to bottom the curves are
 for $q > q_{extr},$ $q\,=\,q_{extr},$ and $q\,<\,q_{extr}.$ The
 double root corresponds to the extremal event horizon.}
\end{figure}

To solve equation $f(r)\,=\,0$  let us introduce a new `radial'
coordinate $x$ and a new parameter $q$ by means of
$ r\,=\, M x$ and $Q^{2}\, = \,q^{2} M^{2},$ respectively. 
Subsequently introducing  a new
 unknown function ${\tilde W}$ 
 \begin{equation}
 x = - {4 q^{2}\over 4 {\tilde W}\,-\,q^{2}},
                                           \label{subs}
 \end{equation}
 one  arrives at
 \begin{equation}
 \exp ({\tilde W}) {\tilde W} = -{q^{2}\over 4}\exp ({q^{2}/4}) .
                                           \label{wz}
 \end{equation}
 As the result of simple manipulations one can relate the exact 
 location of the event horizon $\rp\,(=M x_{+})$ and 
 the inner horizon $r_{-}\,(=M x_{-})$ 
 with the 
 the real branches of the Lambert functions
 \begin{equation}
 x_{+}\,=\,- {4 q^{2}\over 4 W_{0}( -{q^{2}\over 4} \exp (q^{2}/4)) - q^{2}} ,
                                            \label{xpl}
 \end{equation}
 and
 \begin{equation}
 x_{-}\,=\,- {4 q^{2}\over 4 W_{-1}( -{q^{2}\over4} \exp (q^{2}/4)) - q^{2}}.
                                           \label{xmin}
 \end{equation}
 The horizons $r_{+}$ and
 $r_{-}$ for
 \begin{equation}
q_{extr}\,=\,2 w_{0}^{1/2},
                                          \label{qextr}
 \end{equation}
 merge at 
 \begin{equation}
 x_{extr}\,=\,{4 w_{0} \over 1 + w_{0}}
                                          \label{xextr}
 \end{equation}
into a degenerate event horizon.
 Numerically one has 
 \begin{equation}
 x_{extr}\,=\,0.871,\hspace{1cm}{\rm and}\hspace{1cm}q_{extr}\,=\,1.056.
 \end{equation}
 The three types of solutions therefore are: 
the regular black hole with the inner and event horizons for $q <q_{extr},$
the extremal black hole for $q = q_{extr},$ and the regular configuration for 
$q > q_{extr}.$ At large distances as well as for small charges the geometry 
(\ref{lineel}) resembles that of the
Reissner-Nordstr\"om  with one notable distinction: for 
$q > 1$ the Reissner-Nordstr\"om solution describes naked singularity whereas
the regular geometry (\ref{lineel}) could be interpreted as a particle like solution.
Qualitative behaviour of $f(r)$ is displayed in Fig.~\ref{figgg}.

To investigate the near-horizon geometry of the regular magnetic black
hole further let us postpone analyses of the solution (\ref{lineel})
for a while and consider the line element of the form
\begin{equation}
ds^{2}\,=\,\frac{1}{hy^{2}}\left( -dt^{2}\,+\,dy^{2}\right)\,+\,\ro^{2}\dO.
                                      \label{brdeformed}
\end{equation}
Other  useful representations 
obtained through a redefinition of coordinates according to formulas
\begin{eqnarray}
h^{1/2} t&=&e^{\tau} \coth \chi ,\nonumber \\ 
h^{1/2} y&=&e^{\tau} \sinh^{-1}\chi
                                     \label{set1} 
\end{eqnarray}
and
\begin{eqnarray}
\sinh^{2} \chi &=& {\cal R}h -1,\nonumber \\
\tau h&=& {\cal T}
                                    \label{set2}
\end{eqnarray}
are
\begin{equation}
ds^{2}\,=\,\frac{1}{h}\left( -\sinh^{2} \chi dt^{2}\,+\,d\chi^{2}\right)\,+\,
\ro^{2}\dO
                                   \label{br1}
\end{equation}
and
\begin{equation}
ds^{2}\,=\,-\left({\cal R}^{2} h\,-\,1\right) d{\cal T}^{2}\,+
\,\frac{d{\cal R}^{2}}{{\cal R}^{2} h\,-\,1}\,+\,
\ro^{2}\dO,
                                  \label{br2}
\end{equation}
respectively.
Topologically it is ${\rm AdS}_{2}\times {\rm S}^{2},$ i.e., a direct
product of the two-dimensional anti-de Sitter geometry and two-sphere
of constant curvature. The curvature scalar for the line element
(\ref{brdeformed}) is simply a sum of the curvatures of
the subspaces ${\rm AdS}_{2}$ and ${\rm S}^{2}:$
\begin{equation}
R\,=\,K_{{\rm AdS}_{2}}\,+\,K_{{\rm S}^{2}},
\end{equation}
where
$K_{{\rm AdS}_{2}}\,=\,-2 h$ and
$K_{{\rm S}^{2}}\,=\,2/\ro^{2}.$

We intend to solve the coupled system of equations 
of the nonlinear electrodynamics and general relativity.
For the line element~(\ref{brdeformed}) the Einstein tensor is given by
\begin{equation}
G_{a}^{b}\,=\,{\rm diag}\left[ -\frac{1}{\ro^{2}},\,
-\frac{1}{\ro^{2}},\,h,\,h\right],
                                 \label{Eins}
\end{equation}
whereas the electromagnetic tensor compatible with assumed 
symmetries is simply
\begin{equation}
{\tilde F}\,=\,Q\sin\theta d\theta \wedge d\phi ,
                                 \label{Faraday}  
\end{equation}
where $Q$ is, as before,  the magnetic charge.
It could be easily verified that 
\begin{equation}
{ F}\,=\,F_{ab}F^{ab}\,=\,\frac{2Q^{2}}{\ro^{4}}>0.
                                 \label{FSq}
\end{equation}
As the stress-energy tensor of the nonlinear electrodynamics is
\begin{equation}
T_{a}^{b}\,=\,\frac{1}{4\pi}\left(\frac{d\LF}{dF}F_{ac}F^{bc}\,-
\,\frac{1}{4}\delta_{a}^{b}\LF\right),
                                \label{stressT}
\end{equation}
the Einstein fiels equations reduce to two independent equations:
\begin{equation}
\frac{1}{\ro^{2}}\,=\,\frac{1}{2}\LF
                                 \label{rE1}
\end{equation}
and
\begin{equation}
h\,=\,2\left( \frac{d\LF}{dF}F_{23}F^{23}\,-\,\frac{1}{4}\LF\right).
                                 \label{rE2}
\end{equation}

Now we shall select the particular form of the Lagrangian.
Let us choose $\LF $ as given by (\ref{lagr})
with
 \begin{equation}
 a\,=\,\frac{|Q|^{3/2}}{2\alpha},
 \end{equation}
where $\alpha$ is some positive parameter of dimension of length.
Introducing a new coordinate $x$ and a new parameter 
$q$  by means of the formulas 
$\ro\,=\,\alpha x$ and
$Q\,=\,\alpha q,$  Eqs.~(\ref{rE1}) 
and (\ref{rE2}), after some manipulations, could be rewritten as
\begin{equation}
\frac{q^{2}}{x^{2}}\cosh^{-2}\left(\frac{q^{2}}{2 x} \right)\,=\,1
                                    \label{rE1a}
\end{equation}
and
\begin{equation}
h\,=\,\frac{q^{2}}{x^{4}\alpha^{2}}\cosh^{-2}\left(\frac{q^{2}}{2x} \right)\,-\,
\frac{q^{4}}{2x^{5}\alpha^{2}}\frac{\sinh\left(\frac{q^{2}}{2x} \right)}
{\cosh^{3}\left(\frac{q^{2}}{2x} \right)}.
                                 \label{rE2a}
\end{equation}
One can relegate hyperbolic functions combining Eqs.~(\ref{rE1a}) and (\ref{rE2a})
\begin{equation}
\frac{1}{x^{2}}\,-\,\frac{q^{2}}{2x}\left( 1\,-
\,\frac{x^{2}}{q^{2}}\right)^{1/2}\,=\,h\alpha^{2}.
                               \label{br}
\end{equation}

\begin{figure}[th]
 \includegraphics[height=4cm]{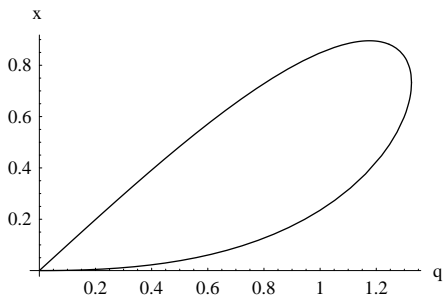}
 \caption{\label{figggg} 
 Curvature radii of $S^{2}$ (solutions of Eq.~(\ref{rE1a})) plotted against
 parameter $q.$}
\end{figure}

In general the problem should be treated numerically: for a particular
choice of $q$ the equation (\ref{rE1a}) gives concrete values of $x$
and hence $h.$ First, observe that depending on $q,$ Eq.~(\ref{rE1a})
has two, one or has no solutions at all. Numerical calculations
indicate that for $q_{c}\,=\,1.325$ there is only one solution
at $x\,=\,0.735,$ whereas for $q\,<\,q_{c}$ Eq.~(\ref{rE1a}) has two
solutions. Moreover, it should be noted  that there is a particular
combination of $q$ and $x$  expressible in terms of the Lambert
functions satisfying Eq.~(\ref{rE1a}). Indeed, it could be easily
checked that  $q\,=\,q_{extr}$ and $x\,=\,x_{extr}$ (given respectively by
Eqs.~(\ref{qextr}) and (\ref{xextr})) comprises such a solution with
$h$ given by
\begin{equation}
h\,=\,\frac{1}{32\alpha^{2}}\frac{(1\,+\,w_{0})^{3}}{w_{0}^{2}}.
                               \label{gauss1}
\end{equation}
Results of numerical calculations are displayed in Fig.~\ref{figggg}.

Although Bertotti-Robinson line element~\cite{Bertotti,Robinson} is a special case of
(\ref{brdeformed}) with $h\,=\,1/\rp^{2}$ it does not belong to a
family of solutions of Eq.~(\ref{rE1a}). It could be easily shown that
vanishing of the trace of the stress-energy tensor (\ref{stressT}) is, by (\ref{FSq}),
possible only for Lagrangians
$
{\cal L}(F)\propto F,
$
and, consequently, ${\cal L}(F)$ in the form given by (\ref{lagr})
does not admit  solutions of the Bertotti-Robinson type.

Now, let us return to the black hole
solution.
The extremal black hole is described by a line element (\ref{lineel}) 
with
\begin{equation}
f(r)\,=\,1\,-\,\frac{2 M}{r}\left[ 1 \,-\,
\tanh\left(\frac{2 M w_{0}}{r} \right)\right].
                                   \label{elex}
\end{equation}
It could be easily shown that $x_{+}$ given by (\ref{xextr}) is the degenerate
event horizon.
In order to investigate the geometry of the vicinity of the 
 event horizon and to obtain uniform approximation
we introduce new coordinates ${\tilde t}$ and $y:$
\begin{equation}
{\tilde t}\,=\,\frac{t}{\varepsilon}, \hspace{1cm} 
r\,=\,\rp\,+\,\frac{\varepsilon}{A\,y},
\end{equation}
where
\begin{equation}
A\,=\,\frac{1}{32 M^{2}}\frac{\left(1\,+
\,w_{0}\right)^{3}}{w_{0}^{2}}.
                                 \label{A1} 
\end{equation}
Expanding the function $f(r)$ in terms of $\varepsilon,$ 
retaining quadratic terms
and subsequently taking the limit $\varepsilon \,=\,0$
we obtain 
\begin{equation}
ds^{2}\,=\,\frac{1}{A\,y^{2}}\left( -dt^{2}\,+\,dy^{2}\right)\,
+\,\rp^{2}\dO.
                          \label{brdeformed2}
\end{equation}
Since $A^{-1}\,>\,\rp^{2},$ the line element does not belong to the
Bertotti-Robinson class, as expected. This result could be easily
anticipated as the stress-energy tensor of the nonlinear
electromagnetic field has nonvanishing trace at the event horizon.
It should be noted that putting $\alpha\,=\,M$ in~(\ref{brdeformed}) with~(\ref{gauss1})
one obtains~(\ref{brdeformed2}). 

It is evident that the procedure of constructing the near horizon geometry 
is insensitive to the nature of the central region of the black hole. One expects,
therefore,
that similar consideration could be carried out for the electrically
charged Ay\'on--Beato and
Garc\'{\i}a solution as modified by Burinskii and Hildebrandt. 

Finally let us return to the problem of boundary conditions and restrict
discussion to the solutions describing black holes. Assuming that the
exact location of the event horizon is known, the boundary condition
$m(\rp)\,=\,\rp/2$ gives
\begin{eqnarray}
m(r)&=&\frac{\rp}{2}\,+\,\frac{|Q|^{3/2}}{2a}\tanh\left(\frac{a|Q|^{1/2}}{\rp} \right)
\nonumber \\
&&-\,\frac{|Q|^{3/2}}{2a}\tanh\left(\frac{a|Q|^{1/2}}{r} \right).
\end{eqnarray}
Denoting first two (constant) terms in the right hand side of the above equation by $M_{H}$
and demanding $a\,=\,|Q|^{3/2}/2M_{H}$ results in the equation which relates
$\rp,$ $Q,$ and $M_{H}$ 
\begin{equation}
1\,-\,\frac{2M_{H}}{\rp}\left(1\,-\,\tanh\frac{Q^{2}}{2 M_{H} \rp} \right)\,=\,0,
\end{equation}
and the line element (\ref{lineel}) with (\ref{fr}). The location of the inner
horizon is given by Eq.~(\ref{xmin}) and the near horizon geometry is, of course,
identical to the one considered previously.






\end{document}